# Magnetic control of an exciton–polariton condensate in a van der Waals magnet

Heng Zhang[1,†], Niloufar Nilforoushan[1,6,†], Christian Weidgans[1,†], Tobias Inzenhofer[1], Marlene Liebich[1], Josef Riepl[1], Julian Hirschmann[2,3], Imke Gronwald[1], Kseniia Mosina[4], Zdeněk Sofer[4], Ritaj Tyagi[5], Jan Wilhelm[5], Fabian Mooshammer[5], Florian Dirnberger[2,3,*], and Rupert Huber[1,5]

*[1]Department of Physics, University of Regensburg, 93040 Regensburg, Germany*

*[2]Zentrum für QuantumEngineering (ZQE), Technical University of Munich, Garching, Germany*

*[3]Department of Physics, Technical University of Munich, Munich, Germany*

*[4]Department of Inorganic Chemistry, University of Chemistry and Technology Prague, 166 28 Prague 6, Czech Republic*

*[5]Regensburg Center for Ultrafast Nanoscopy, University of Regensburg, 93040 Regensburg, Germany*

*[6]present address: Université Paris Cité, CNRS, LMPQ, Paris, France*

*[†]These authors contributed equally*

**corresponding author: f.dirnberger@tum.de*

**Quasiparticle condensates are among the most spectacular solid-state manifestations of quantum physics. Coupling macroscopic real-space wavefunctions to additional degrees of freedom, such as the electron spin, would add valuable control knobs for quantum applications. While creating spin-carrying superconducting condensates has attracted enormous attention, man-made condensates of light-matter hybrids known as exciton-polaritons have lacked an analogous spin-based perspective. Here we open a new door by demonstrating magnetically tunable exciton-polariton condensation in the van der Waals magnet CrSBr. Under photoexcitation, CrSBr micro-wires embedded in an optical cavity show the hallmarks of polariton condensation: a dramatic increase of the emission intensity from an excited laterally confined polariton state by multiple orders of magnitude, spectral narrowing of the emission line, and a continuous shift of the peak energy. Interferometry evidences an increase in spatial and temporal coherence. Owing to the strong coupling between the spin order and excitonic correlation, the energy of the condensate can be tuned by up to 10.5 meV by an external magnetic field of only 2 Tesla. Our results establish CrSBr microcavities as a powerful platform for exploring magnetic control of polariton condensates and mark a significant step toward spin-controlled coherent quantum light sources.**

## Introduction

Non-equilibrium Bose-Einstein condensates in solids have formed a unique platform for exploring macroscopic quantum phenomena[1,2], ranging from superfluid propagation[3] to nonlinear interactions[4]. Condensates of hybrid light-matter quasiparticles[5,6], called exciton-polaritons, are particularly promising owing to their combination of photonic coherence and excitonic interactions, which makes them a candidate for quantum communication and cryptography applications[7,8]. The key challenge, however, lies in controlling the coherent condensate emission. So far, most polariton condensates have been realized in conventional semiconductor microcavities where quasiparticle interactions are weak and tunability typically requires a modification of the structure. Quantum materials that mediate interactions of exciton-polaritons with other quasiparticles, such as phonons or magnons, could provide a compelling path forward. While proximitizing superconducting condensates with spin-ordered materials has led to new functionalities[9-12], exciton-polariton condensation in spin-ordered systems remains largely unexplored.

Magnetic van der Waals (vdW) crystals[13-15] have recently emerged as an attractive class of quantum materials supporting profound interactions between excitons, photons, phonons, and magnons[16-21]. Particularly the air-stable layered A-type antiferromagnet CrSBr (Fig. 1a) stands out for its tightly bound, quasi-one-dimensional excitons which are polarized along the crystallographic *b*-axis with giant oscillator strength[22]. The most prominent exciton resonance of bulk CrSBr occurs near an energy of 1.36 eV. At the surface, this resonance is shifted to typically 1.34 eV (ref. [23]). Thus far, strong coupling of excitons to vertically confined photon modes in bulk crystals has been explored, where new magneto-optic correlations[24,25], hyperbolic effects[26], anisotropy[27], and most recently, signatures of bulk polariton condensation[28] emerge. Exfoliated flakes of CrSBr form microscopic wires which are strongly elongated in the *a*-axis, effectively realizing a slab waveguide geometry with additional lateral photon confinement.

Here, we demonstrate the condensation of exciton-polaritons in such CrSBr microwires. Under ultrafast optical excitation, our microwires show the fingerprint of polariton condensation – that is, a dramatic increase in the photoluminescence (PL) intensity accompanied by linewidth narrowing and a pronounced energy shift. Optical interferometry of the PL emission reveals a clear increase in its spatial and temporal coherence. Because a fully spin-polarized electronic manifold is underlying this condensate, its energy can be magnetically changed by up to 10.5 meV. We find that optical pumping of surface exciton states most efficiently drives condensation, pointing to an ultrafast excitation transfer via resonant magnon and phonon pathways. These spin-coupled exciton-polariton condensates open new avenues for designing and controlling waveguided coherent emission by external magnetic fields, magnons, and spin fluctuations. They may hence be used to interface semiconductor optics and future photonic quantum technologies with spin-ordered matter.

**Magnetic microwire polaritons**

To achieve strong coupling between bulk excitons and a single photon mode confined in the vertical direction, few-nm CrSBr crystals are centred between two planar cavity mirrors using PMMA spacers (Methods). In an optical microscope, we identify few-layer CrSBr crystals that are elongated along the crystallographic $a$-axis with typical lengths of tens of μm but exhibit a small width of approximately 1 μm in the $b$-direction (see Fig. 1b for an exemplary microcavity). Such structures can act as microwires that spatially confine light like slab waveguides. The tightest confinement is given in the vertical ($c$-axis) direction. The resulting strong coupling with the bulk exciton resonance and the emergent polaritonic effects are thus overall well-described by numerical calculations assuming infinite lateral crystal extensions (see Extended Data Fig. 1). A narrow width of the flake along the $b$-axis, however, causes additional quantization[29] (Fig. 1b, inset), which is less relevant along the $a$-axis owing to the extended wire length in this direction. This situation can be qualitatively captured by modelling the polariton as a particle in an elongated 2D rectangular potential well (see Methods). The model predicts the confinement to cause a sequence of discrete quantized polariton states along the $b$-axis with energy splittings on the order of tens of meV (Fig. 1c). Since the polaritons are barely restricted along the $a$-axis, the dispersion of the polariton branches remains essentially parabolic along this direction (Fig. 1d).

We first characterized the polariton PL of our microwires in the antiferromagnetic phase (lattice temperature, $T_{\mathrm{L}} = 5$ K), where the spins are aligned antiparallel between adjacent layers (Fig. 1a). Orienting CrSBr wires parallel or perpendicular to the entrance slit of the spectrometer allows us to image the polaritonic band dispersion along the two orthogonal angles $\theta_a$ and $\theta_b$ (Fig. 1c,d, insets). Light emitted at an angle $\theta_{a,b}$ originates from exciton-polaritons with energy $E$ and in-plane momentum $k_{a,b} = (E/\hbar c)\sin\theta_{a,b}$ (refs. [1,5,30]). As an example, we discuss the light emission of a CrSBr cavity with a width of $W \approx 1.25$ µm, and a length of $L \approx 40$ µm (labelled cavity #1, Extended Data Fig. 1a), after optical excitation (pulse duration, 240 fs; repetition rate, 85 kHz) at a photon energy of $\hbar\omega_{\text{pump}} = 1.771$ eV.

Resolving the PL spectra along the $b$-axis for -25° < $\theta_b$ < 25°, we observe discrete features consistent with the lateral quantization of polaritons in narrow optical waveguides (Fig. 1e). For better visibility, the PL intensity of weakly emissive excited states (marked by a white dashed box) is upscaled. Several quantized modes can be clearly recognized, including the ground state ($Q_0$), the first excited state ($Q_1$) and the second excited state ($Q_2$) at energies of 1.248 eV, 1.265 eV and 1.286 eV, respectively. Conversely, the PL intensity along the $a$-direction displays a series of dispersive, parabolic polariton modes exhibiting minima at those energies (Fig. 1f). This is in line with the negligible quantization along the $a$-axis. Details of the dispersion vary between samples[31-33] (Extended Data Figs. 2 and 3) owing to differences in thickness and width of the wires, yet they all exhibit a continuous parabolic dispersion along the (long) $a$-axis and discrete energy states along the (short) $b$-axis, consistent with Fig. 1c,d.

For a quantitative analysis of the exciton-photon coupling strength, we implemented a global fit to the quantized and parabolically dispersed polariton states by a coupled oscillator model (Fig. 1e,f, white dots and black solid lines; see Methods for details). The good agreement with the experimental data indicates that these modes correspond to the lower polariton emission of a 9 nm-thick (~10 layers) CrSBr crystal[34]. The interaction between vertically confined photons and excitons reaches the strong-coupling regime with a vacuum Rabi splitting of $2\hbar\Omega_{\text{R}} = 200$ meV, which matches recent observations of polaritonic effects in CrSBr[24,25].

Besides the lower polaritons, we observe two isolated resonances at $\hbar\omega \approx 1.312$ eV and 1.325 eV (Fig. 1f). Although they are coupled to the same cavity mode, the absence of notable dispersion indicates that these resonances remain in the weak coupling regime. Considering the different dielectric environment and strain in our cavity, these features can tentatively be ascribed to the recently identified surface excitons of CrSBr[23]. This assignment is further supported by a much smaller magnetic-field-induced shift of surface excitons compared to bulk excitons (Extended Data Fig. 4). Surface states at energies well below the bulk exciton resonance open a pathway for near-resonant excitation of polaritons that favours condensation, as we show below.

**Fingerprinting exciton-polariton condensation**

In a polariton condensate, final-state stimulation leads to a macroscopic population of a single quantum state, which should manifest in a dramatic PL increase with increasing polariton density[5,32,35]. To test this criterion, we measure the emission of cavity #1 as a function of pump fluence $\Phi$. We first tune our pump laser into resonance with the dispersionless surface exciton band at $\hbar\omega_{\mathrm{pump}} = 1.325$ eV. The resulting PL spectra (Fig. 2a) unveil a strongly nonlinear fluence dependence of the intensity from $Q_1$. Under near-resonant pumping, $Q_1$ appears slightly blue shifted with respect to the peak observed in Fig. 1c owing to the different many-body effects induced by unbound electron-hole pairs and a dense polariton gas, respectively. The spectrally integrated peak emission exhibits a prototypical threshold-like scaling with the pump fluence (Fig. 2b). Above the pump threshold $\Phi_{\mathrm{th}} \approx 30\ \mu\mathrm{J/cm^2}$, a moderate increase of the fluence by a mere factor of 3 causes a dramatic enhancement of the PL peak centred at $\hbar\omega = 1.276$ eV by more than two orders of magnitude. In sharp contrast, the polariton emission scales linearly when the cavity is excited off-resonantly at $\hbar\omega_{\mathrm{pump}} = 1.771$ eV (Fig. 2b and Extended Data Fig. 5).

To resolve the nature of the polariton states giving rise to such a strongly nonlinear increase of the PL emission, we acquired angle-resolved PL maps (Fig. 2c,d and Extended Data Fig. 6). The dominant PL peak in Fig. 2a originates from $Q_1$ confined by our microwire in the lateral $b$-direction, which is separated from the pump photon energy by 46 ± 3 meV. Whereas the overall PL emission of $Q_0$ and $Q_1$ is comparable below threshold, $Q_1$ completely dominates the PL signal for $\Phi > \Phi_{\mathrm{th}}$ (Fig. 2d). The abrupt increase of the PL intensity from $Q_1$ is accompanied by a significant decrease of its linewidth, from 19 meV to 8.7 meV, and an energy increase by 5 meV (Fig. 2e,f). These features are unambiguous hallmarks of exciton-polariton condensation in microcavities. Note that the momentum spread of the polariton distribution along the $b$-axis is not expected to narrow upon condensation because $Q_1$ is a spatially confined state.

Final-state stimulation inherently synchronizes the phase of all particles contributing to the condensate. The resulting spontaneous macroscopic spatial and temporal coherence should persist as long as the condensate lives. We prove this key aspect by measuring the first-order correlation function of the polariton PL emission of cavity #1 in a Michelson interferometer setup[5,32]. The PL emission is first split evenly into the two arms of the interferometer, before one of the beams is inverted centro-symmetrically by a retroreflector. The two beams are then combined by a lens and overlapped in the real-space image plane. Temporal overlap is set by an optical delay line. Figures 3a,b show the spatially overlapped PL images at a delay time of $t$ = 0 fs below and above the threshold, respectively. While signatures of coherence are absent below $\Phi_{\mathrm{th}}$, interference fringes evidence spatially coherent emission for $\Phi > \Phi_{\mathrm{th}}$.

Sweeping $t$ allows us to measure the temporal coherence of the condensate emission (Fig. 3c-e). We quantify the coherence time by extracting the visibility of the interference fringes, defined as $\nu = (I_{\mathrm{max}} - I_{\mathrm{min}})/(I_{\mathrm{max}} + I_{\mathrm{min}})$, as a function of $t$ (Fig. 3f,g). Here $I_{\mathrm{max}}$ ($I_{\mathrm{min}}$) represents the intensity of a local maximum (neighbouring minimum) of the fringes. The data are well described by a Gaussian profile with a full width at half maximum (FWHM) coherence time of 427 ± 7 fs. This value corresponds to a linewidth of ~4 meV, which is compatible with the width observed in Fig. 2e if the finite resolution of the spectrometer is considered (Methods). Most importantly, the measured coherence time exceeds the spontaneous polariton lifetime by one order of magnitude (Methods), unambiguously proving that the exciton-polaritons in CrSBr undergo condensation. The combination of a superlinear intensity scaling, an energy blueshift of the emission peak, and enhanced coherence times signifies that the condensation process emerges spontaneously. Similar signatures are also observed in other comparable microcavities, even at elevated temperatures above 20 K, and in PL spectra dispersed along the $a$-axis (Extended Data Fig. 7).

**Unique aspects of polariton condensation in CrSBr**

Several unique aspects about our polariton condensates deserve special mention. First, we observe condensates only under near-resonant excitation of surface excitons, which lie energetically between the bulk exciton resonance and its polaritons (Fig. 2b and Extended Data Fig. 5). Pumping at 1.771 eV, in contrast, yields no condensation despite the much larger excitation densities created by strong B-exciton absorption[21,23]. The origin of this difference likely lies in the different relaxation pathways following high-energy excitation. Populating polariton states requires a sequence of scattering events, each constrained by energy-momentum matching. The phase space of such relaxation into the narrow condensate region is strongly reduced, and competing channels, such as redistribution of excitation energy into other excitonic states, radiative loss before relaxation, or scattering into a broad incoherent reservoir dominate.

Another notable observation is that polariton condensates are formed in excited rather than ground states of our microwires. In two of our samples (Fig. 2 and Extended Data Fig. 3), the energy difference between pump photons and the excited-state condensate emission closely matches the energy of an optical magnon mode of CrSBr (~45 meV), which has recently been evidenced in neutron scattering[36] and far-infrared spectroscopy[37]. Besides ultrafast excitation of the magnetic order and magnons[38], emission of multiple $A_g^1$ phonons, which are known to couple strongly with excitons in CrSBr[39,40], or a single $A_g^3$ phonon should also efficiently populate the condensate (see Extended Data Fig. 8). To understand the migration of electronic excitation, we calculated the lattice relaxation of the lowest energy exciton state in CrSBr. Density functional theory (DFT) calculations reveal that electrons localize on the Cr atoms while holes localize on the S atoms. This charge distribution induces a structural distortion that increases the interlayer separation, coupling efficiently to the interlayer breathing $A_g^1$ phonon mode (see Extended Data Fig. 9). This strongly supports efficient excitation energy transfer via scattering with $A_g^1$ phonons. Together, these results point to a consistent picture in which resonant phonon- and magnon-assisted scattering channels provide an efficient scattering pathway for transferring excitation energy from surface to bulk states. We suggest that wire-to-wire variations in confinement can tune polariton states into the right energy window for supporting resonant enhancement of these processes, thereby enabling the formation of non-equilibrium condensates in excited microwire states.

Lastly, the fluence dependence of polariton energies observed in our study is qualitatively different from previous reports. So far, either a decrease of polariton energies[24,41], or a competition between opposite shifts of polariton energies[28] was noted with increasing excitation fluence. We, however, observe a monotonic increase of polariton energies (Fig. 2f). Circumventing other relaxation pathways, it thus seems that near-resonant excitation at kHz repetition rates isolates a contribution to the polariton energy that shifts its states towards higher energies. This is indeed typically ascribed to repulsive Coulomb interactions between excitons[1,7,28,41,42] or saturable absorption effects[43,44]. In CrSBr, however, additional contributions may arise from the coupling of excitons to spin order: A joint theory-experiment study has recently shown that magnon-induced magnetic disorder can sizably reduce the oscillator strength of excitons in CrSBr[23,26]. It is therefore important to note that when excitons are strongly coupled to photons, a reduction in oscillator strength decreases the Rabi gap and consequently increases the lower polariton energy. This effect is particularly pronounced for lower polaritons with sizable photon fraction, like those studied in our microwires. Hence, we suggest polariton energy shifts in CrSBr may involve contributions, not only from exciton-exciton repulsion, but also from the (unavoidable) photoexcitation of magnons[45,46] and the resulting renormalization of the Rabi gap.

**Magnetic field control of polariton condensates**

In any case, the exciton-polariton modes in CrSBr emerge from a manifold of spin-polarized electronic orbitals. Hence, spin order should provide a natural control knob for polariton condensates in CrSBr. To explore this idea, we mount a new microwire (cavity #5) into a magneto-optical cryostat with tunable out-of-plane magnetic fields (Extended Data Fig. 10). Figure 4a shows the angle-resolved PL emission from this microwire containing laterally quantized polariton states $Q_0$, $Q_1$, and $Q_2$, the dispersionless emission from surface excitons ($X_S$), and a defect emission peak at 1.265 eV labelled $X_D$ (ref. [47]).

We first tracked the magnetic field evolution of these laterally confined polariton modes under cw excitation. As the magnetic field increases from 0 T to 2 T, both the exciton-polariton branches and the surface exciton resonance exhibit pronounced redshifts of more than 10 meV, whereas the emission by localized defects remains nearly unaffected. Shape and magnitude of these shifts evidence their origin from a field-driven reconfiguration of the collective spin order rather than spin Zeeman splitting or the magnetic polaron effects observed with condensates in dilute magnetic semiconductors[48,49,50]. Importantly, the parabolic spectral shift saturates above 2 T, consistent with the known antiferromagnetic-to-ferromagnetic transition in CrSBr under out-of-plane magnetic bias. Each confined mode, $Q_0$, $Q_1$, and $Q_2$, inherits a distinct magnetic field shift directly from their respective exciton fraction and the absolute shift of bulk excitons (~18 meV, refs. [21,24,39]; see Extended Data Fig. 4d). Compared to bulk excitons, the shift of surface excitons is notably reduced owing to their different magnetic environment causing a weaker dependence on the external field. All in all, our observations obtained under cw excitation demonstrate that magnetic order directly renormalizes the polaritonic eigenstates.

We now extend this magnetic control to the condensation regime. Like in our other microwires, the quantized polariton mode $Q_1$ exhibits the characteristic superlinear increase in PL intensity at threshold both in the antiferromagnetic ($B$ = 0 T) and ferromagnetic (2 T) phases. This confirms that condensation can be achieved in both the antiferromagnetic and the ferromagnetic configuration. Most strikingly, angle-resolved measurements reveal that the condensed state itself undergoes a field-induced energy shift of up to 10.5 meV. Such large tuning of a macroscopically coherent quantum state mediated by the spin order stands in contrast to previous electric-field-based strategies[51,52,53]. While the latter perturbatively modify the potential experienced by exciton-polaritons, the magnetic field in CrSBr changes magnetic order, symmetry, and ultimately, the nature of the exciton wavefunction from quasi-one-dimensional species to spatially extended three-dimensional wavefunctions as the material transitions from antiferromagnetic to ferromagnetic order[22]. To the best of our knowledge, this is the first control of an exciton–polariton condensate mediated by spin order. Given the uniquely large spin-induced energy shift, this concept opens an extremely promising interface between macroscopic quantum states and spintronics.

## Conclusion

We demonstrate polariton condensation in microwires of the van der Waals antiferromagnet CrSBr. The polaritons are laterally quantized by the finite dimension of the microwire along its short axis. Near-resonant pumping causes polariton condensation in excited quantized states, evidenced by a multi-order increase of polariton PL emission, linewidth narrowing, energy shift and spatio-temporal coherence. Because such condensates are intimately coupled to the underlying spin order, their energy can be shifted by up to 10.5 meV through reconfiguration of the magnetic order. Photoexcitation of surface states appears to enable particularly efficient population of polariton condensates in CrSBr microwires through phonon and magnon scattering pathways. Our results open exciting possibilities for interfacing the macroscopic quantum mechanical wavefunction of tailor-made condensates with spin order, magnetic control, and spintronic concepts for next-generation quantum technologies. Particularly appealing are the prospects of modulating coherent condensate emission on ultrafast timescales by coupling it to coherent, high-frequency magnons. This platform could enable magnetic-memory integration and efficient microwave-to-optical transduction for quantum information purposes, where microwave photons couple to condensate emission via magnons.

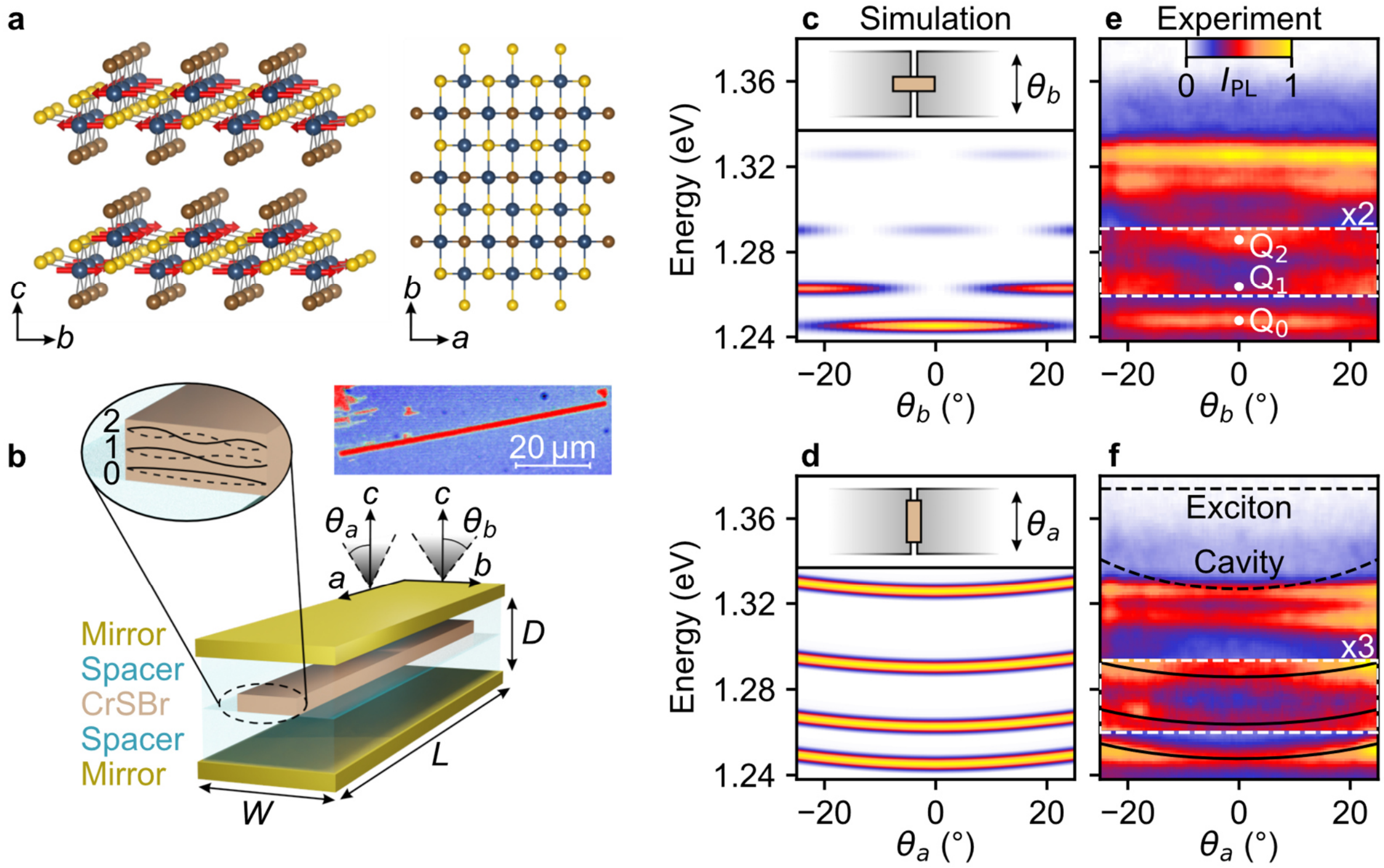

**Figure 1 | Exciton-polaritons in a CrSBr cavity wire. a**, Crystal structure of CrSBr viewed along the *a*- and *c*-axes (dark spheres, Cr; yellow spheres, S; brown spheres, Br). Red arrows indicate the magnetic order in the antiferromagnetic phase. **b**, Schematic of the CrSBr microcavity structure. The zoom-in indicates the spatial profile of the laterally quantized polariton modes. An exemplary false-colour micrograph of a CrSBr microcavity is shown at the top right corner. **c**, Simulated polariton dispersion along the short axis of the microwire obtained by modelling polaritons as particles in an elongated 2D rectangular potential well. Inset: In the experiment, we resolve this dispersion when the *b*-axis of the microwire is aligned parallel to the entrance slit of the spectrometer. **d**, Simulated dispersion along the long axis of the microwire, equivalent to the experimental case where the *a*-axis of our microcavity is aligned parallel to the entrance slit of the spectrometer (inset). **e**, Angle-resolved normalized PL of cavity #1 dispersed along the *b*-axis, and **f**, along the *a*-axis measured using ultrafast photoexcitation (1.771 eV; $\Phi$ = 94 μJ/cm$^2$; pulse duration, 240 fs). PL signals in the white rectangular boxes are enhanced to increase visibility. The white dots in **e** indicate the ground state ($Q_0$), and the first ($Q_1$) and second ($Q_2$) excited states obtained from a coupled oscillator model (see main text). The black dashed lines in **f** show the dispersions of uncoupled exciton and cavity photon modes. The solid black lines display the theoretically fitted lower polariton branches from the coupled oscillator model.

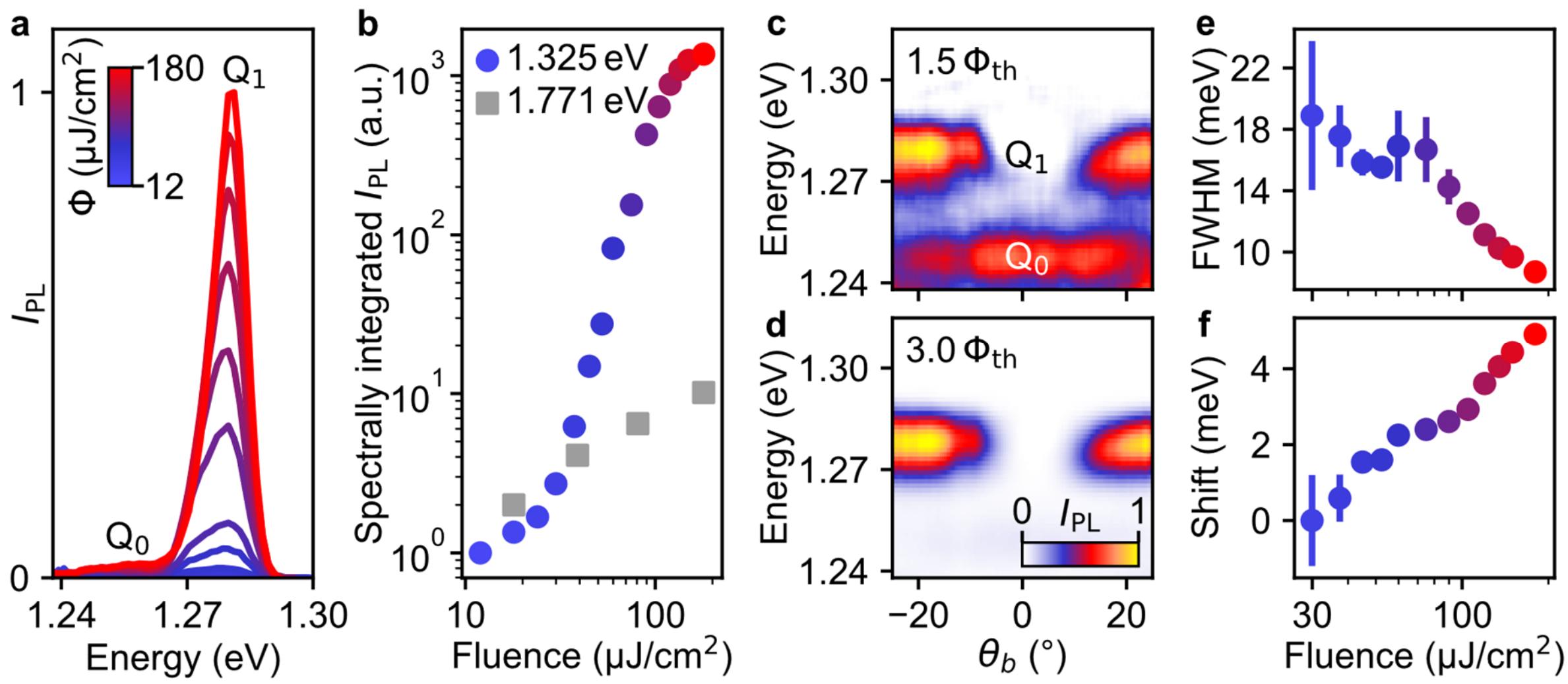

**Figure 2 | Nonlinear photoluminescence response and onset of exciton-polariton condensation.** **a**, Fluence-dependent normalized PL spectra of cavity #1 measured for a pump photon energy of 1.325 eV at 5 K. **b**, Fluence-dependent PL intensity for the state $Q_1$ (blue-to-red dots) at 1.276 eV in **a**. The grey squares represent the PL intensity at the same peak measured under off-resonant excitation using a photon energy of 1.771 eV. **c-d**, Normalized angle-resolved PL spectra at pump fluences of 45 μJ/cm$^2$ and 90 μJ/cm$^2$. **e-f**, Linewidths and energy shifts of the PL peak at 1.276 eV in **a** as a function of pump fluence.

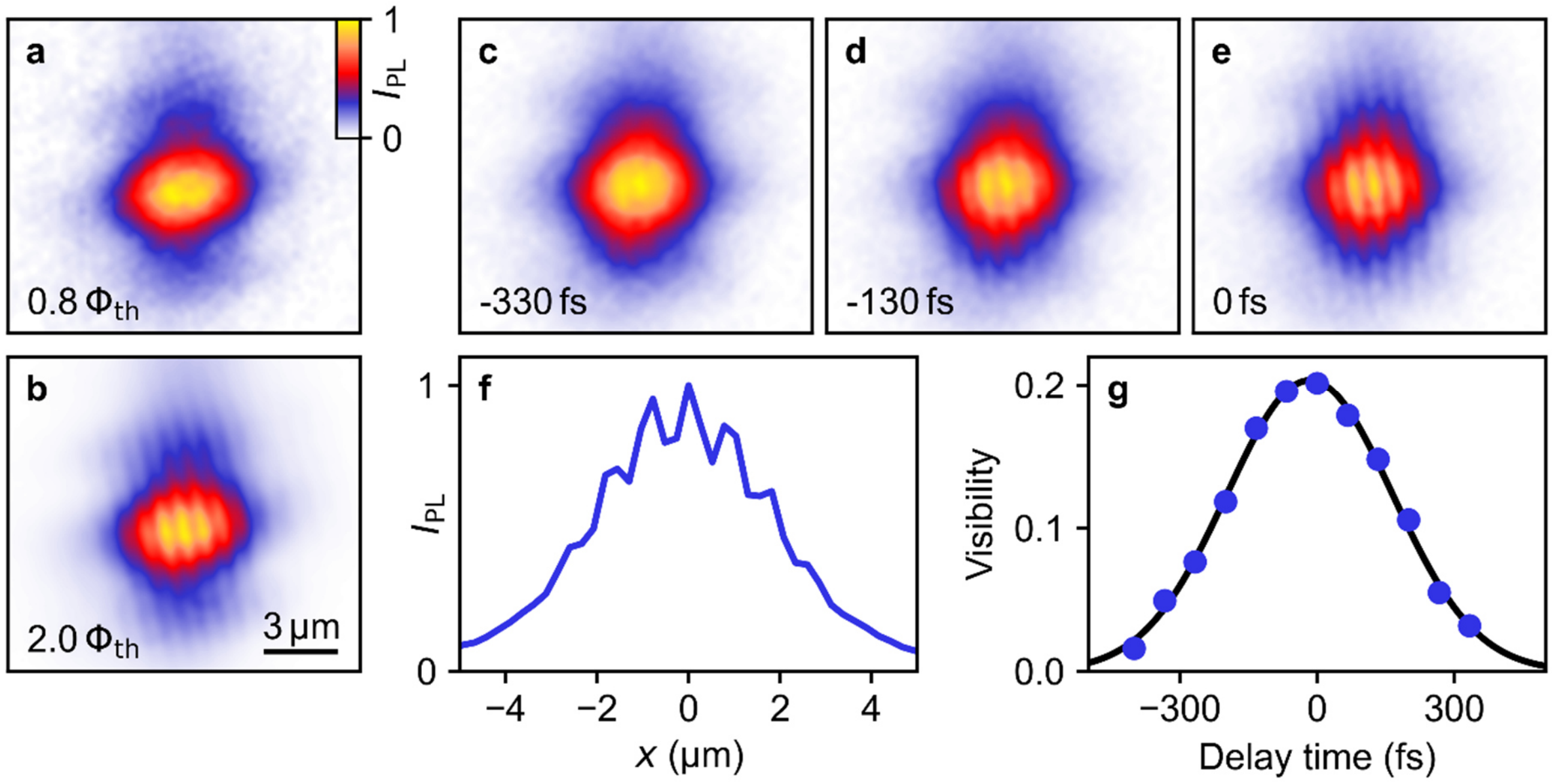

**Figure 3 | Spatial and temporal coherence of the exciton-polariton condensate.** Zero-delay spatial interference pattern of the exciton-polariton PL of cavity #1 at 5 K and a pump fluence of **a**, $0.8\Phi_{th}$ and **b**, $2.0\Phi_{th}$, where $\Phi_{th}$ = 30 µJ/cm$^2$. **c-e**, Spatial interferogram of the exciton-polariton PL emission measured at delay times of $t$ = -330 fs, -130 fs and 0 fs at a pump fluence of $2.0\Phi_{th}$. **f**, Interference fringes averaged along the direction perpendicular to the fringes for $t$ = 0 fs and a pump fluence of $2.0\Phi_{th}$. **g**, Visibility of the interference pattern in **f** as a function of delay time $t$ at a pump fluence of $2.0\Phi_{th}$.

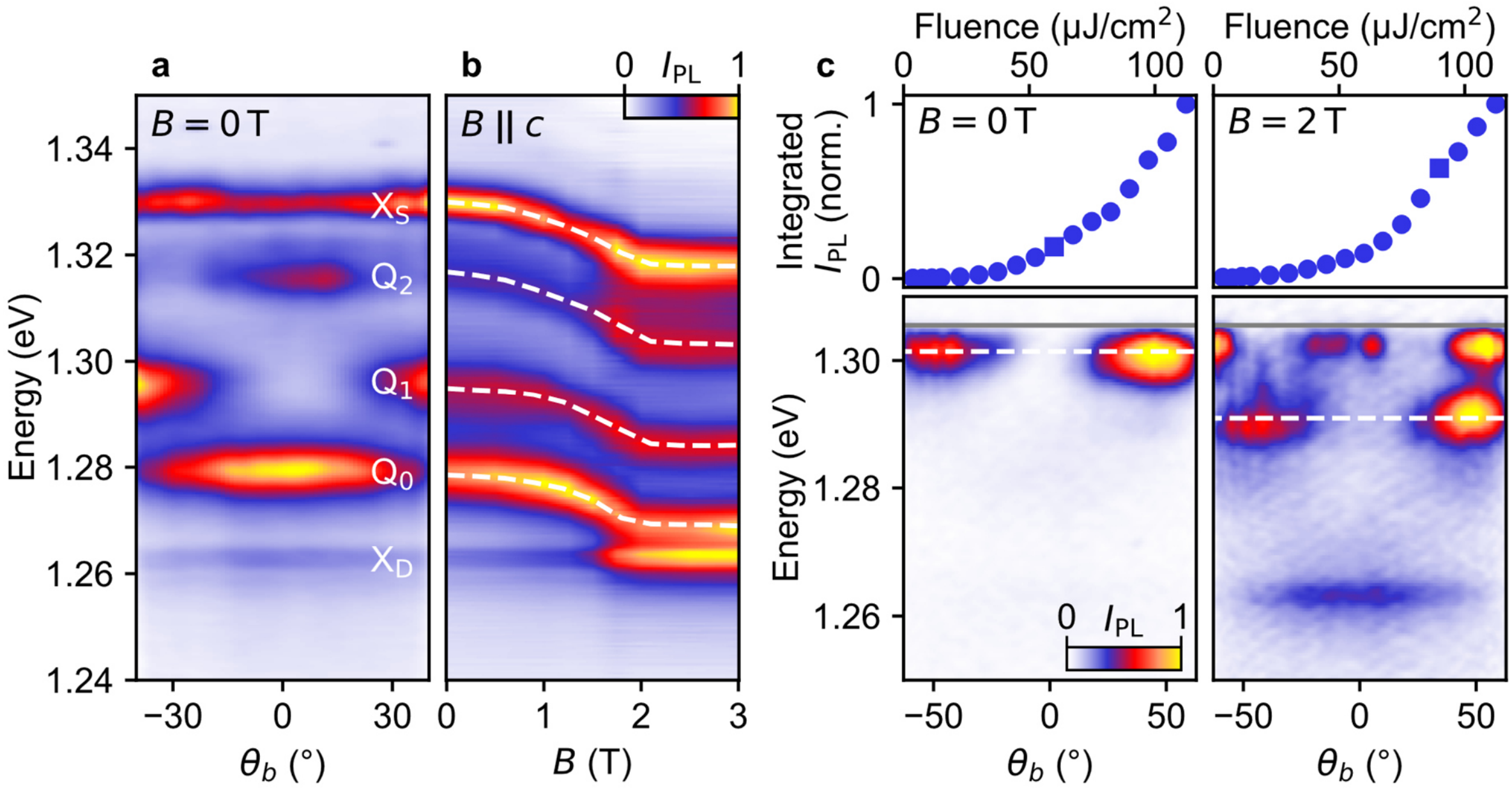

**Figure 4 | Magnetic control of an exciton-polariton condensate. a**, Angle-resolved PL spectra along the *b*-axis for *B* = 0 T, measured at 10 K under excitation by a cw laser with a photon energy of 1.907 eV. $Q_0$, $Q_1$ and $Q_2$ represent the quantized polaritonic ground state, the first excited state and the second excited state, respectively. $X_S$ and $X_D$ are the surface exciton state and a defect state. **b**, Normalized, angle-integrated PL intensity as a function of the magnetic field *B* oriented along the *c*-axis. The dashed lines serve as guides to the eye. **c**, Top panels: integrated PL intensity of state $Q_1$ as a function of pump fluence at *B* = 0 T (left) and *B* = 2 T (right). The data are taken at 10 K under pulsed excitation with a pump photon energy of 1.325 eV. The blue squares indicate the pump fluences used for the corresponding angle-resolved PL spectra in the bottom panels. Bottom panels: angle-resolved PL spectra along the *b*-axis for *B* = 0 T (left) and *B* = 2 T (right) with a pump fluence of 60 μJ/cm² and 90 μJ/cm², respectively. White dashed lines denote the central energy of state $Q_1$, revealing a magnetically induced energy shift of the condensate of up to 10.5 meV. Grey solid lines mark the photon energy below which our long-pass spectral filter suppressing the incident pump light allows PL intensity to be collected.

## Methods

**Fabrication of the microwire cavities.** A 200-nm-thick gold mirror was evaporated on a silicon wafer and covered by a spacer layer of PMMA (nominal thickness, 70 nm). Subsequently, CrSBr was transferred onto the substrate by micro-mechanical exfoliation, capped by another 70 nm PMMA spacer and a semi-transparent, 40-nm-thin gold mirror to complete the microcavity. The schematic layout is illustrated in Fig. 1b.

**Optical spectroscopy and microscopy characterizations.** For the optical measurement, the microcavities are pumped by ultrafast laser pulses (repetition rate: 85 kHz, pulse duration: 240 fs) from an optical parametric amplifier (ORPHEUS; Light Conversion) driven by a commercial regenerative Yb:KGW laser amplifier system (PHAROS 10; Light Conversion). The relatively low repetition rate helps us avoid pulse-to-pulse recovery time artefacts and allows sufficient pulse energies for future nonlinear optical experiments. For optical imaging of the sample, we utilize a tungsten-halogen light source. To detect the photoluminescence, we use a spectrometer (Acton SP2500) coupled with a Peltier-cooled silicon charge-coupled device (CCD) camera (PIXIS 256) with a resolution of 1024 × 256 pixels. The width of the entrance slit was kept at several 100 µm to achieve sufficient statistics despite the low repetition rate. This setting leads to a linewidth broadening of typically ~6 meV. Cosmic-ray events on the CCD camera and slight asymmetries in Fourier images were digitally corrected. For the experiments of Figs. 1-3, the sample is mounted in a closed-cycle cryostat and cooled down by liquid helium to 5 K. To perform magnetic control studies (Fig. 4) we placed the microwire samples in a magnet cryostat with a large-aperture optical access and a variable out-of-plane magnetic field of up to 5 T.

Our optical setup allows us to access both the real-space and momentum-space images of the PL from the sample. To obtain the spectrum and real-space image, the emission from the sample is projected onto the imaging plane, i.e., the entrance slit of the spectrometer. To resolve the momentum (angle) distribution of the emission, an additional lens is placed into the beam path, such that the Fourier plane of the objective lens is projected on the entrance slit of the spectrometer.

The coherence measurements are conducted in a Michelson interferometer, with a retroreflector mounted in one of the two interferometric arms to invert vertically and horizontally the spatial coordinates of the real-space image. The relative temporal delay between the two arms is controlled by an optical delay line. The two interference images are overlapped by a lens on the slit of the spectrometer and imaged by the CCD camera.

**Coupled oscillator model.** The exciton-polariton dispersion in Fig. 1f is fitted by the coupled oscillator model[1]:

$$\begin{pmatrix} E_{\mathrm{cav}} & \hbar\,\Omega/2 \\ \hbar\,\Omega/2 & E_{\mathrm{exc}} \end{pmatrix} \begin{pmatrix} \alpha \\ \beta \end{pmatrix} = E \begin{pmatrix} \alpha \\ \beta \end{pmatrix}$$

Here $E_{\mathrm{cav}}$ and $E_{\mathrm{exc}}$ are the energies of the cavity photon and exciton, respectively. $\hbar\Omega$ is the vacuum Rabi splitting, related to the exciton-photon coupling strength $g$ following $\hbar\Omega = 2g$. Moreover, $\alpha$ and $\beta$ are the amplitudes of the eigenvectors. The eigenvalues of the energy can be obtained by diagonalizing the matrix: $E_{\pm}(k_{\parallel}) = \frac{E_{\mathrm{cav}}+E_{\mathrm{exc}}}{2} \pm \frac{1}{2}\sqrt{[E_{\mathrm{cav}} - E_{\mathrm{exc}}]^2 + \hbar^2\Omega^2}$ with $E_{+}$ and $E_{-}$ being upper and lower polariton branches. Exciton and photon fractions (Extended Data Fig. 4) are determined by the amplitude squared of the Hopfield coefficients $X_{k_{\parallel}}$ and $C_{k_{\parallel}}$, respectively. They are given by

$$\left|X_{k_{\parallel}}\right|^2 = \frac{1}{2}\left(1 + \frac{\Delta E(k)}{\sqrt{\Delta E(k_{\parallel})^2 + 4g_0^2}}\right) \text{ and } \left|C_{k_{\parallel}}\right|^2 = \frac{1}{2}\left(1 - \frac{\Delta E(k_{\parallel})}{\sqrt{\Delta E(k_{\parallel})^2 + 4g_0^2}}\right)$$

with $\Delta E = E_{\mathrm{cav}} - E_{\mathrm{exc}}$. Due to both the vertical and lateral quantum confinement of the cavity, the cavity photon mode can be further expressed as[27-28]:

$$E_{\mathrm{cav}}(j, k_x) = \frac{\hbar c k_z}{n_c}\sqrt{1 + \left[\frac{(j+1)\pi}{L_y}\right]^2 \frac{1}{k_z^2} + \left(\frac{k_x}{k_z}\right)^2}$$

Here, $k_z = \pi/L_z$, is the wavevector along the vertical direction of the cavity with $L_z$ being the effective length of the cavity. $k_x$ and $k_y$ are the lateral wavevectors along the extended direction (*a*-direction in our case), and along the confined direction (*b*-direction), respectively. For each quantized mode $j(0,1,2\,...)$ along *b* direction, $k_y = \pi(j+1)/L_y$ with $L_y$ being the width of the cavity.

**'Particle in a box' simulation.** To obtain the dispersions shown in Fig. 1c,d, we solve the two-dimensional Schrödinger equation using the finite difference method implemented in an open-source code available under https://github.com/LaurentNevou/Q_Schrodinger2D_demo. The CrSBr flake is modelled as a rectangular potential well, whose dimensions are chosen according to the size of the sample. The barrier is set to 10 eV to mimic an infinite potential well. Consequently, the effective mass is the only free parameter of the model. The best agreement with the experimental data is obtained for effective masses $m_a = m_b = 4.8 \times 10^{-5} m_0$ along the $a$-axis $(m_a)$ and $b$-axis $(m_b)$with the free-electron mass $m_0$. For each eigenenergy, the obtained wavefunctions are converted to momentum-space using a fast Fourier transform (FFT). Finally, a Gaussian broadening in energy is applied for better comparability with the experiment.

**Computational details of the density function theory calculations**

Electronic structure calculations were performed using the CP2K suite (version 2025.2) with the Quickstep (GPW) method[54]. To accurately capture the electronic properties of the system, we employed the HSE06 screened hybrid functional[55], incorporating D3 dispersion corrections with Becke-Johnson (BJ) damping[56] to properly account for the long-range van der Waals interactions between the CrSBr layers. The TZVP-MOLOPT-PBE basis set was used in combination with Goedecker-Teter-Hutter (GTH) pseudopotentials. To maintain computational efficiency while treating the exchange correlation, the Auxiliary Density Matrix Method (ADMM) was utilized with a corresponding double-zeta basis set optimized for MOLOPT-GTH basis. The plane-wave cutoff of 800 Ry, and the self-consistent field (SCF) convergence was set to $10^{-8}$ Hartree. Structural optimizations of the excited-state geometries (excitons) were conducted using the Maximum Overlap Method (MOM)[57] to maintain the desired orbital occupation. All calculations were restricted to the **Γ**-point of the Brillouin zone. The two-dimensional CrSBr layers were modelled using a 3×3 supercell of unit cell with in-plane dimensions of 4.74 Å × 3.55 Å with ferromagnetic intralayer coupling and antiferromagnetic interlayer coupling. To minimize artificial periodic interactions along the z-axis, a vacuum spacing was included, resulting in total cell heights of 45 Å for the bilayer and 90 Å for the four-layer systems. Input and Output files of the CP2K calculation are available on Nomad, doi:10.17172/NOMAD/2026.03.03-1

**Data availability**

The data sets generated during and/or analysed during the current study are available from the corresponding author upon reasonable request.

**Acknowledgements**

The work at the University of Regensburg was supported by the Deutsche Forschungsgemeinschaft (DFG, German Research Foundation) through research grants HU1598/8 and SFB 1277, project-ID 314695032. N.N. acknowledges financial support by the Alexander von Humboldt Foundation. F.D. and J.H. acknowledge funding from the Emmy Noether Program (Project-ID 534078167). Z.S. and K.M. were supported by project LUAUS25268 from Ministry of Education Youth and Sports (MEYS) and by ERC-CZ program (project LL2101) from (MEYS). Z.S. was supported by the Advanced Multiscale Materials for Key Enabling Technologies project, supported by the Ministry of Education, Youth, and Sports of the Czech Republic. Project No. CZ.02.01.01/00/22_008/0004558, co-funded by the European Union. F.M. acknowledges funding from the Elite Network of Bavaria (ENB) and funding by the Bavarian State Ministry of Science and the Arts (StMWK). R.T. and J.W. were supported by the DFG via the Emmy Noether Programme (project number 503985532). The authors gratefully acknowledge the computing time on the high-performance computer Otus at the NHR Center Paderborn Center for Parallel Computing, which is jointly supported by the Federal Ministry of Research, Technology and Space and the state governments participating in the National High-Performance Computing (NHR) joint funding program (www.nhr-verein.de/en/our-partners).

**Author contributions**

F.D. and R.H. conceived the study. H.Z., N.N. and C.W. built the optical setup. C.W., T.I. and J.R. designed the magnetic-field-biased experiments. H.Z., N.N., C.W., T.I. and M.L. carried out the experiments. H.Z., N.N., J.H. and F.D. computed the polariton dispersions. F.M. performed the quantization simulations. K.M. and Z.S. grew the CrSBr flakes. I.G. prepared the cavity samples. R.T. and J.W. carried out the DFT calculations. H.Z., C.W., T.I., J.H., and F.D. prepared the figures. All authors analysed the data, discussed the results, and contributed to the writing of the manuscript.

**Competing interests**

The authors declare no competing interests.

**Correspondence and requests for materials** should be addressed to F. Dirnberger.

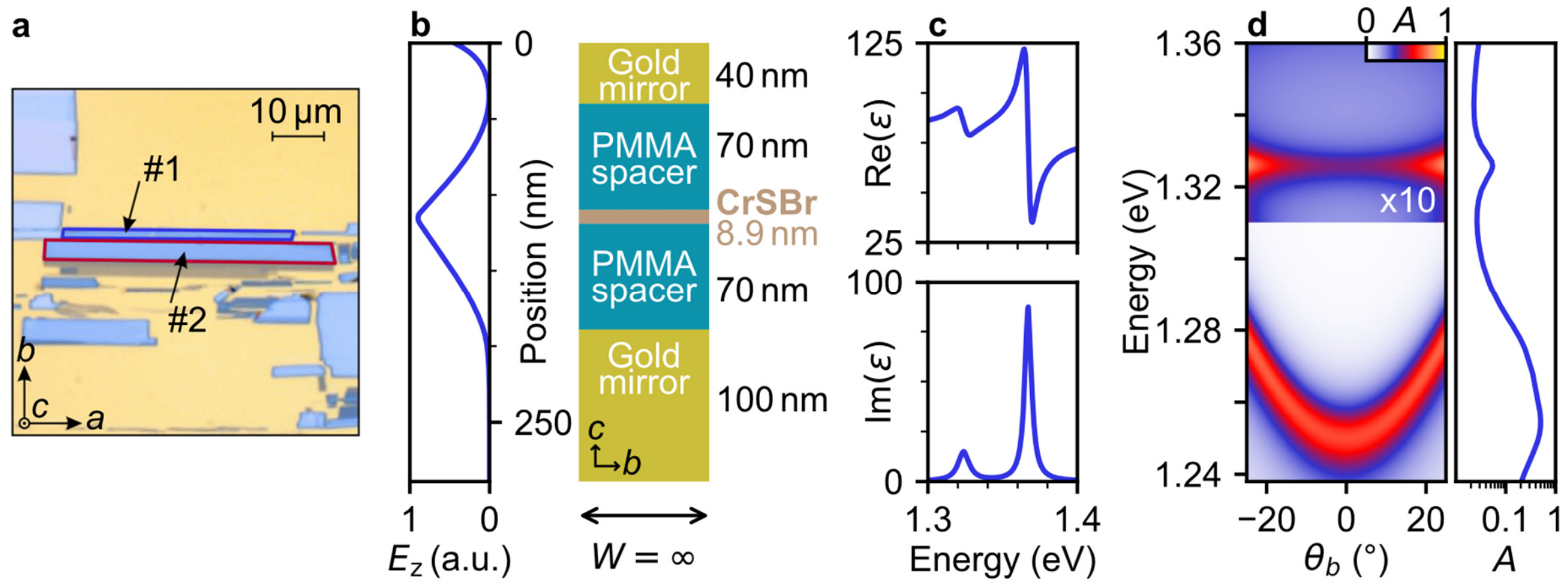

**Extended Data Figure 1 | Characterization of a typical CrSBr cavity. a,** False-colour optical micrograph of cavity #1 and #2. **b,** Schematic of the CrSBr cavity (right) and the electric field distribution along the *c*-direction (left). **c,** Real and imaginary parts of the CrSBr dielectric function used for numerical calculations. A Lorentz oscillator model accounts for contributions from bulk excitons, at 1.36 eV, and from surface excitons, at 1.33 eV (ref. [23]). **d,** Numerically calculated absorption of cavity #1 assuming infinite lateral extensions of the flake.

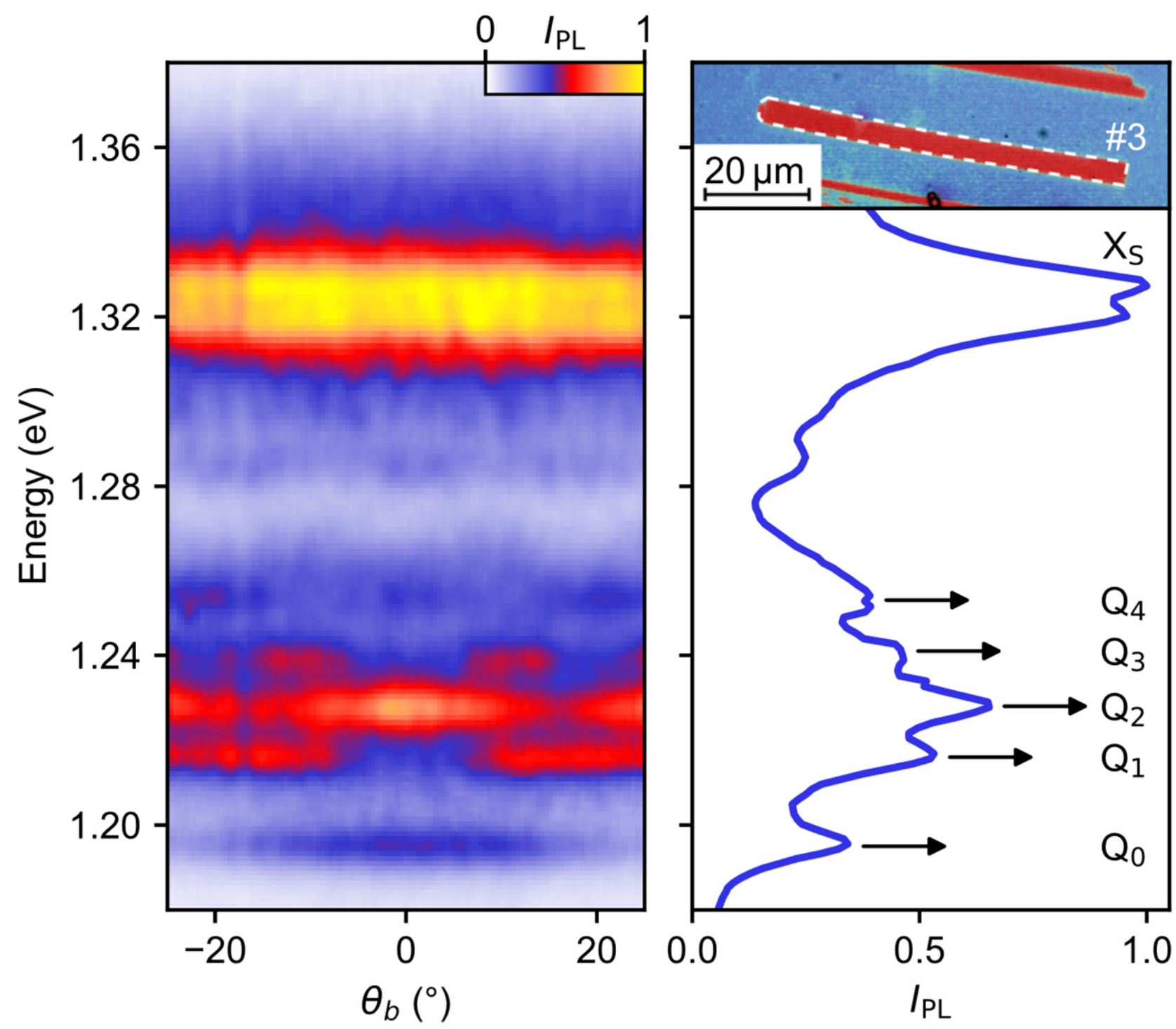

**Extended Data Figure 2 | Photoluminescence of cavity #3.** Left: Angle-resolved normalized PL along the *b*-axis, pumped by 1.771 eV laser excitation with a pump fluence of 300 μJ/cm$^2$. Right: Normalized PL spectrum. Inset: False-colour optical micrograph of cavity #3. In the energy range between 1.19 eV and 1.26 eV, quantized states of $Q_0$-$Q_4$ are well resolved. This cavity serves to illustrate the reproducibility of the polariton confinement and mode quantization.

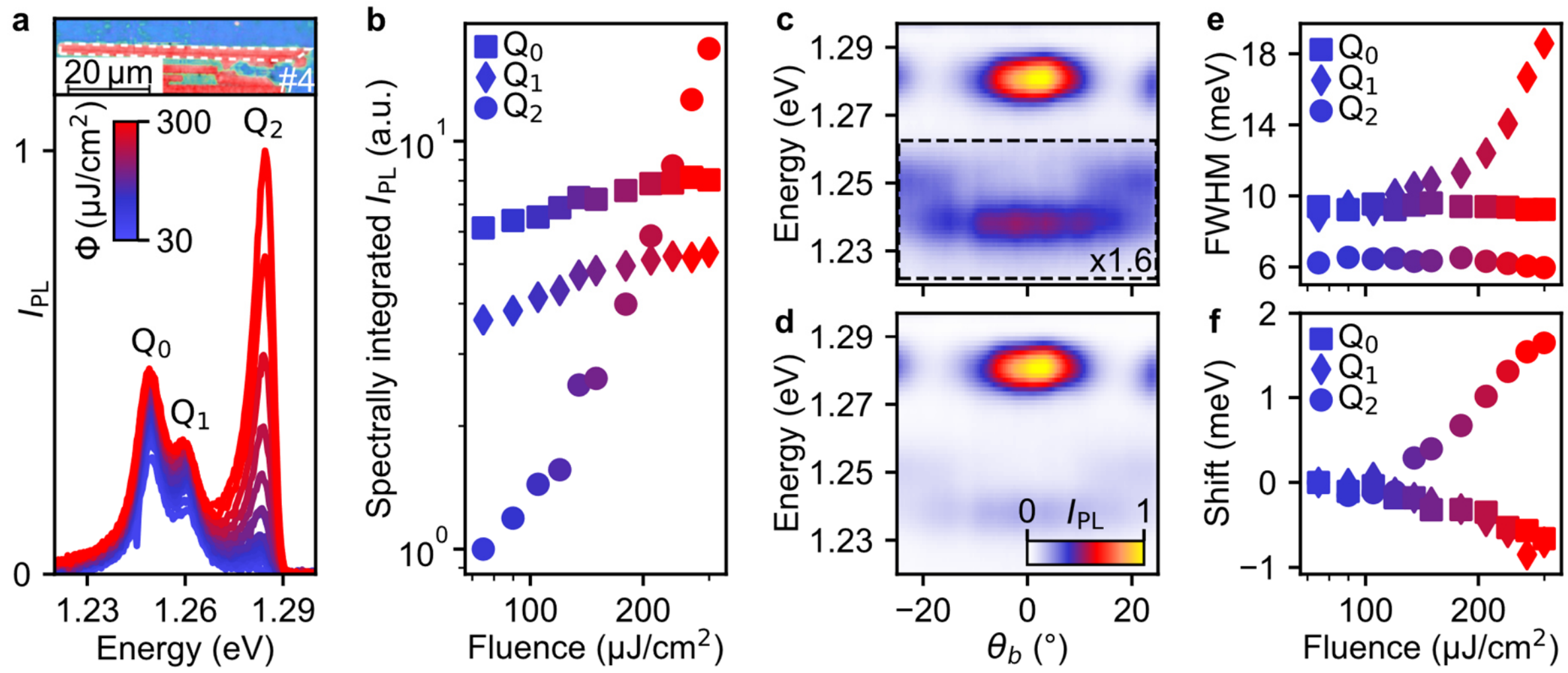

**Extended Data Figure 3 | Nonlinear photoluminescence response and onset of exciton-polariton condensation for cavity #4 pumped by 1.325 eV laser excitation at 5 K. a,** Pump fluence-dependent normalized PL spectra. Inset: False-colour optical micrograph of cavity #4. **b,** Pump fluence-dependent PL intensity of the peaks $Q_0$, $Q_1$, $Q_2$, as labelled in **a**. **c-d,** Angle-resolved PL spectra normalized to the maximum intensity at energies of 1.28 eV, for pump fluences of 150 µJ/cm$^2$ and 300 µJ/cm$^2$ respectively. The PL signal in the black rectangular box in **c** is enhanced to increase the visibility of the $Q_0$ and $Q_1$ states. **e-f,** Linewidths and energy shifts of the PL peaks $Q_0$, $Q_1$, $Q_2$ as a function of pump fluence.

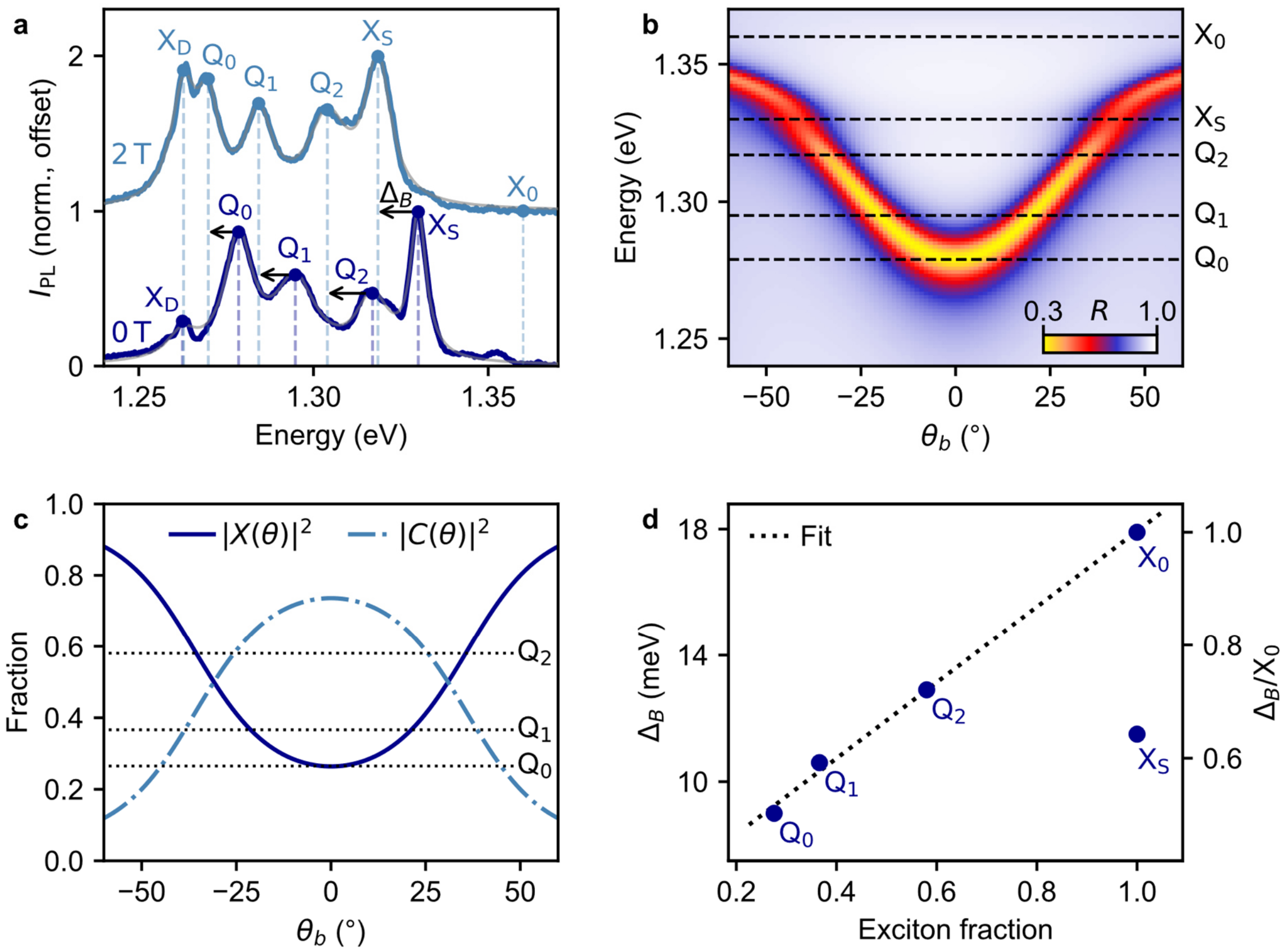

**Extended Data Figure 4 | Distinct magnetic field response of bulk exciton-polaritons and surface excitons.** **a**, Angle-integrated, normalized PL spectra of cavity #5 (see Fig. 4 in the main text) measured at 10 K for $B$ = 0 T and $B$ = 2 T under cw excitation at a photon energy of 1.91 eV. Labels indicate the laterally confined bulk exciton-polariton modes, $Q_0$, $Q_1$ and $Q_2$, and the surface exciton ($X_S$) resonances. Magnetic-field-induced shifts ($\Delta_B$) of each resonance are indicated by arrows. The peak at 1.263 eV ($X_D$) is nearly unaffected by the $B$ field and ascribed to defects[47]. The energy of the bulk exciton resonance ($X_0$) deduced from simulations and used in the Hopfield model is also marked. **b,** Calculated angle-resolved reflectance of a vertical microcavity without lateral confinement. Experimentally measured energy positions of $Q_0$, $Q_1$, $Q_2$, and $X_S$ are indicated along with $X_0$ by dashed lines. **c,** Exciton fraction (squared Hopfield coefficient $X$) of exciton-polaritons and corresponding photon fraction derived by fitting the Hopfield model to the calculated reflectance of **b. d,** Energy shift, $\Delta_B$, as a function of exciton fraction demonstrating the distinct magnetic field response of bulk exciton-polaritons and surface excitons. $X_0$ shift adapted from ref. [24].

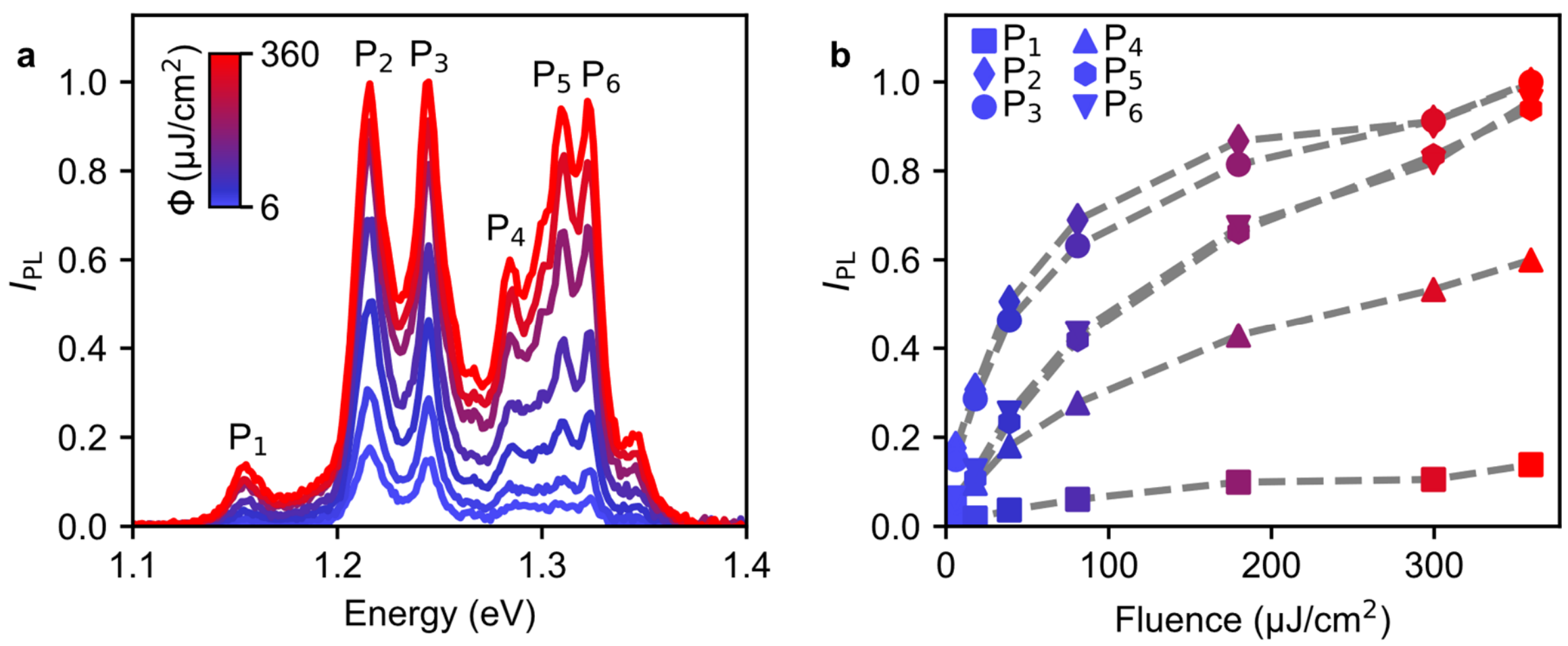

**Extended Data Figure 5 | PL spectra of cavity #1 pumped by 1.771 eV laser excitation at 5 K. a,** Fluence-dependent PL spectrum. $P_3$ corresponds to $Q_0$ of cavity #1 in the main text, while $P_1$ and $P_2$ are the residual signals from cavity #2. **b,** PL intensity for spectrum peaks $P_1$-$P_6$ labelled in **a** as a function of pump fluence.

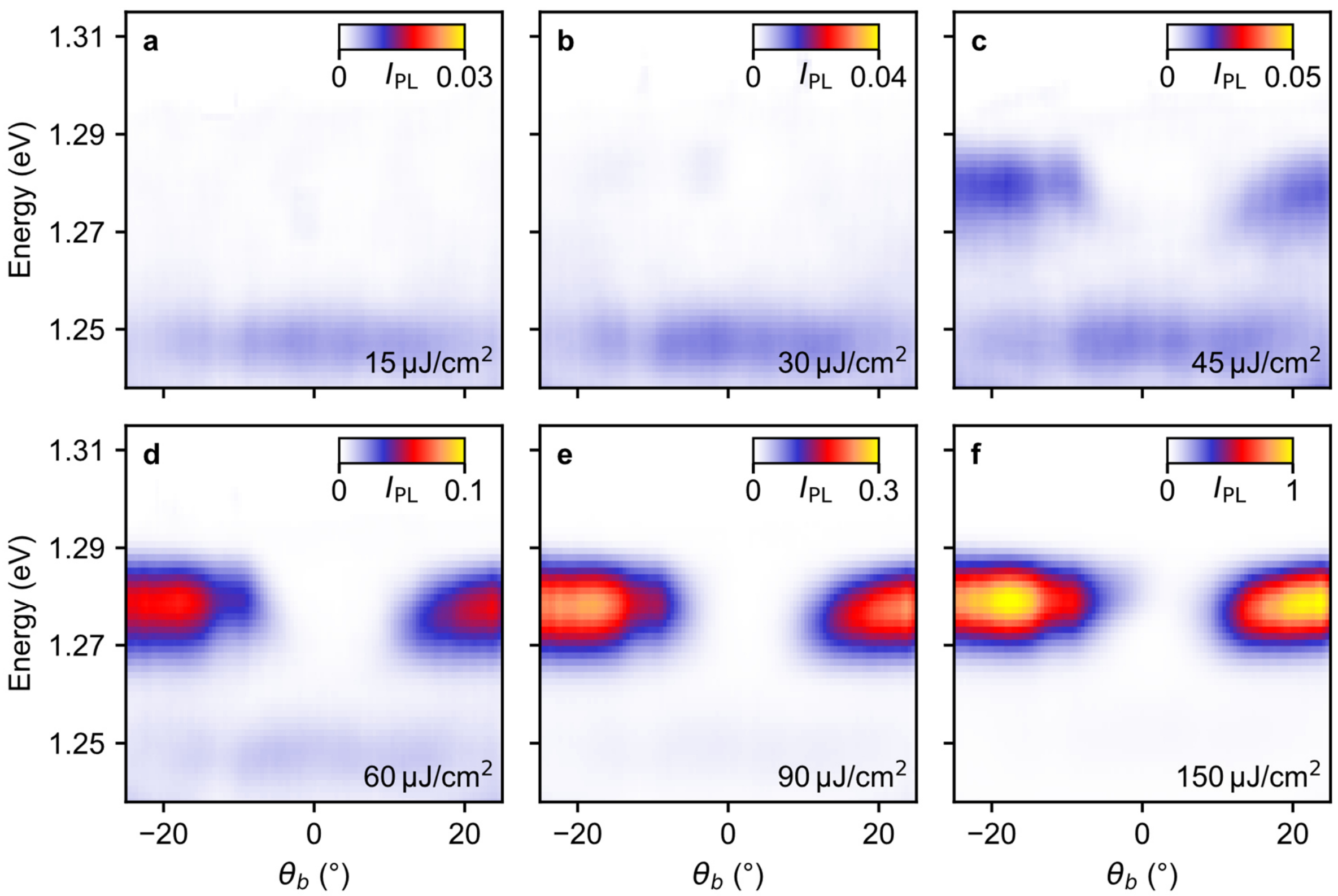

**Extended Data Figure 6 | Angle-resolved PL spectrum at different pump fluences.** Fluence-dependent angle-resolved PL image for cavity #1 at 5 K, pumped by a femtosecond laser centred at a photon energy of 1.325 eV and pump fluences of **a,** 15 μJ/cm$^2$, **b,** 30 μJ/cm$^2$, **c,** 45 μJ/cm$^2$, **d,** 60 μJ/cm$^2$, **e,** 90 μJ/cm$^2$ and **f,** 150 μJ/cm$^2$.

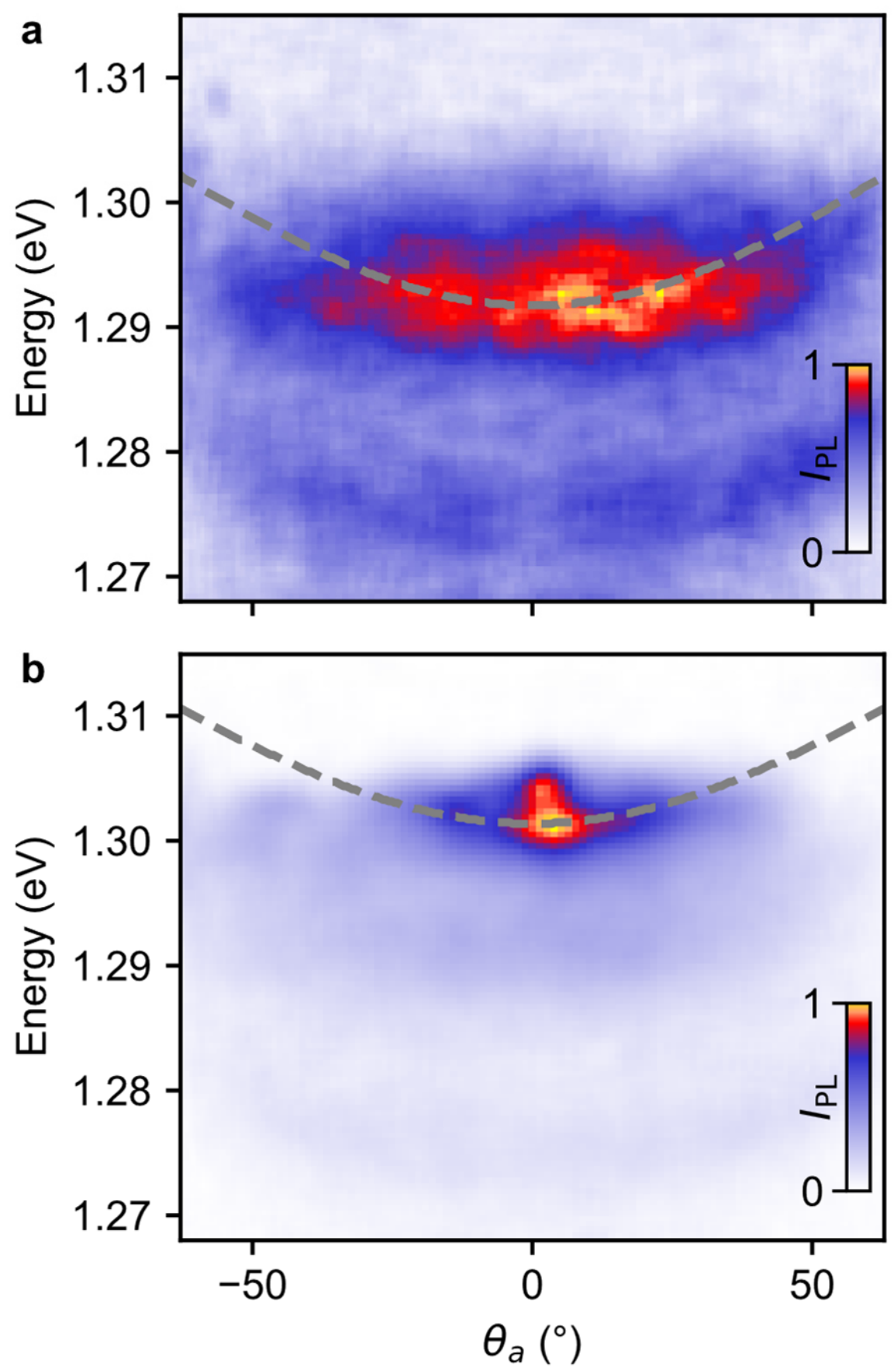

**Extended Data Figure 7 | Exciton-polariton condensation dispersed along the *a*-axis. a,** Angle-resolved PL spectra normalized to the maximum intensity, for cavity #5 below the condensation threshold. **b,** Corresponding above-threshold spectra. Gray dashed lines are guides to the eye indicating the polariton dispersion of state $Q_1$. For low pump fluences, the polariton distribution maps out a parabolic band because the polaritons are effectively unconstrained along the long *a*-axis of the wires. Once the pump fluence exceeds the condensation threshold, only the state at $k = 0$ becomes macroscopically populated, as the condensation leads to a narrowing of the *k*-space distribution.

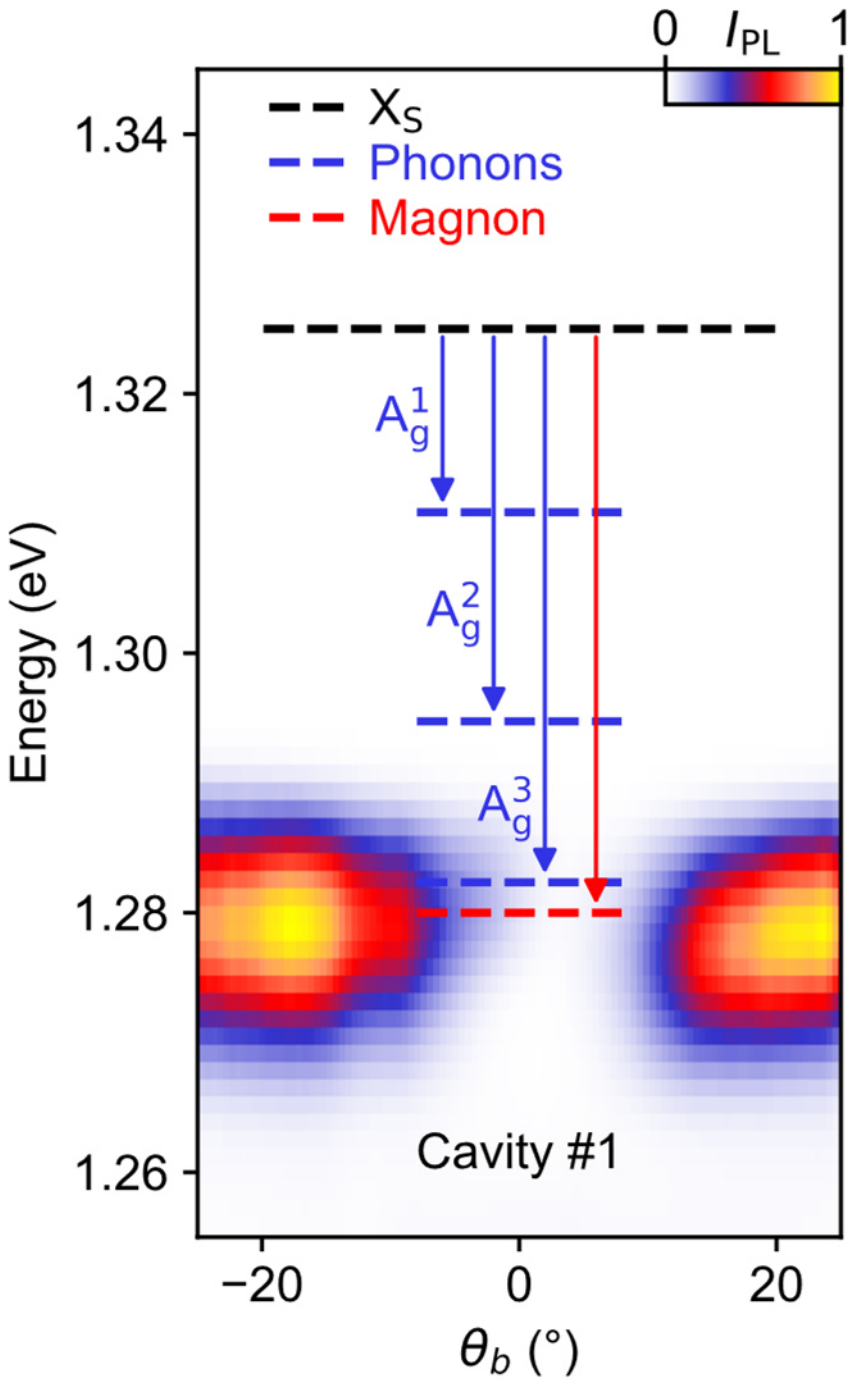

**Extended Data Figure 8 | Quantitative analysis of scattering processes.** Angle-resolved PL spectrum normalized to the maximum intensity, for cavity #1 at $B = 0$ T and $\Phi = 150$ μJ/cm$^2$. The black dashed line indicates the energy of the surface exciton ($X_S$) extracted from Fig. 1e. Blue dashed lines mark the energy after the energy relaxation via a phonon, while the red dashed line indicates relaxation mediated by a magnon. CrSBr hosts three dominant phonon modes labelled $A_g^1$, $A_g^2$ and $A_g^3$.

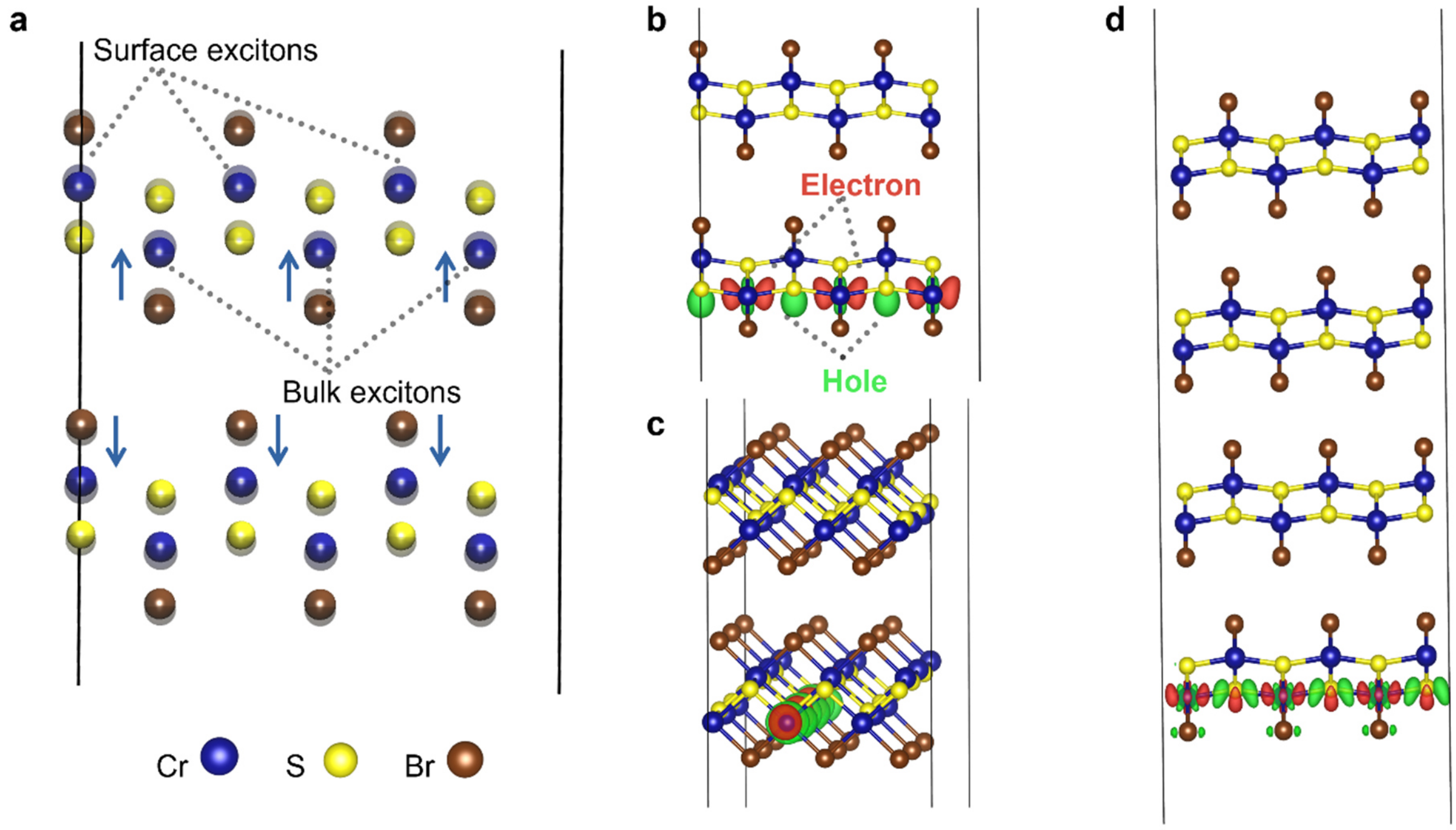

**Extended Data Figure 9 | Excitonic structural distortion and wavefunction localization calculated by DFT.** **a,** Comparison between the pristine (solid) and exciton-optimized (transparent) lattice configurations of a 3×3 supercell of CrSBr bilayer, calculated using the HSE06 functional with D3(BJ) dispersion corrections. The exciton-induced distortion reveals an increase in interlayer separation of ~0.35 Å, corresponding to the $A_g^1$ breathing mode. Isosurface plots of the charge density difference ($\Delta\rho = \rho_{excited} - \rho_{ground}$) are shown in **b,c** for the CrSBr bilayer along different viewing directions, and in **d** for the CrSBr tetralayer. Both densities are evaluated at the excited-state geometry. An isovalue of 0.14 e/Å$^3$ is used. The resulting distributions indicate electron accumulation (red) and depletion (green) with dominant contributions from Cr and S atoms, respectively.

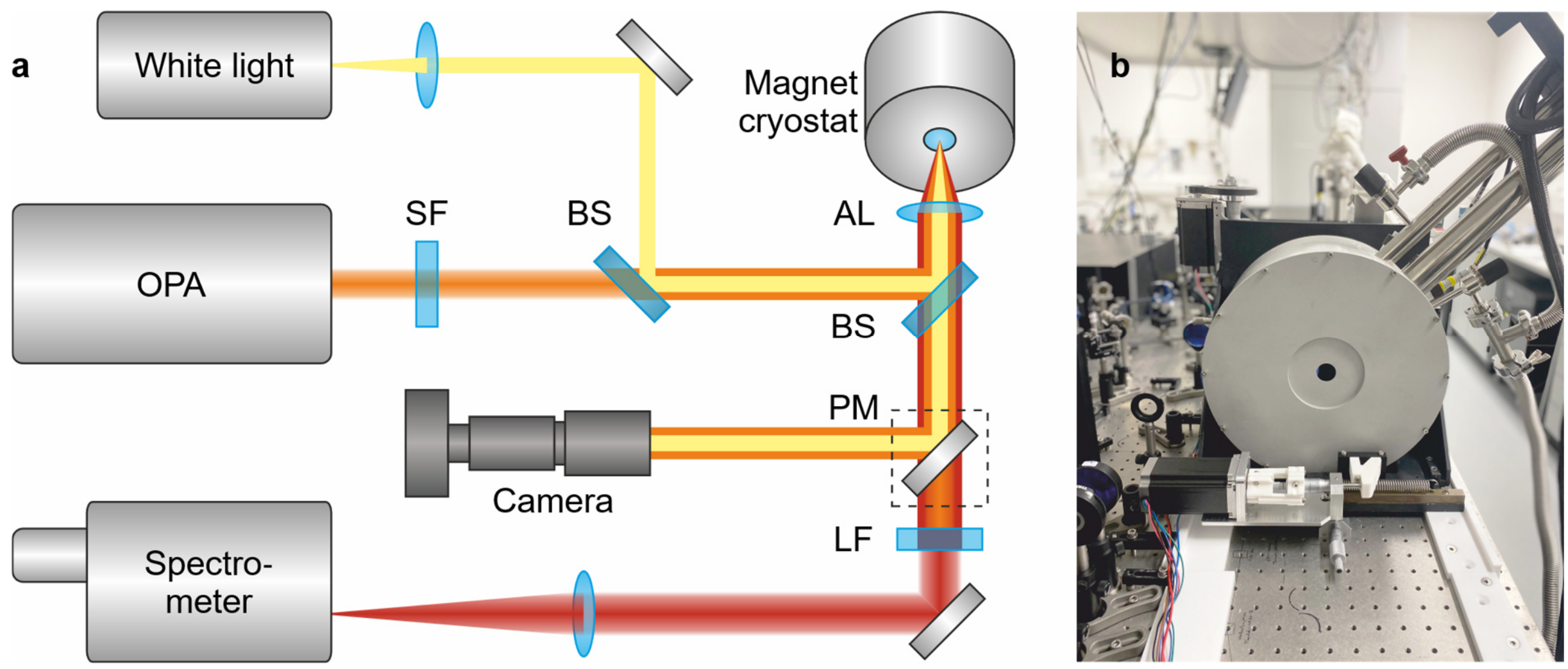

**Extended Data Figure 10 | Schematic sketch of the experimental setup for the magneto-optical measurements. a**, The microcavity is mounted inside a magnet cryostat generating tunable magnetic fields of up to 5 T. For the sample visualisation, white light is directed onto the cavity, while the reflected light is collected by a camera with the help of a temporarily inserted pick-off mirror (PM). For excitation, optical parametric amplifier (OPA) pulses are tightly focused onto the cavity using an aspherical lens (AL) with a high numerical aperture (NA = 0.89). The emitted photoluminescence light is collected by the same lens and guided to the spectrometer. Spectral pre-selection is achieved using a shortpass filter (SF) and a longpass filter (LF). BS, beam splitter. **b**, Photograph of the cryostat, which is mounted on a three-dimensional stage with motorised in-plane sample positioning.